\documentstyle[12pt]{article}

\begin{document}
{\sf \begin{center} \noindent {\Large\bf Lyapunov spectra instability of chaotic dynamo Ricci flows in twisted magnetic flux tubes}\\[2mm]

by \\[0.1cm]

{\sl  L.C. Garcia de Andrade}\\Departamento de F\'{\i}sica
Te\'orica -- IF -- Universidade do Estado do Rio de Janeiro-UERJ\\[-3mm]
Rua S\~ao Francisco Xavier, 524\\[-3mm]
Cep 20550-003, Maracan\~a, Rio de Janeiro, RJ, Brasil\\[-3mm]
Electronic mail address: garcia@dft.if.uerj.br\\[-3mm]
\vspace{1cm} {\bf Abstract}
\end{center}
\paragraph*{}
Previously Casetti, Clementi and Pettini  [\textbf{Phys.Rev.E
\textbf{54},6,(1996)}] have investigated the Lyapunov spectrum of
Hamiltonian flows for several Hamiltonian systems by making use of
the Riemannian geometry. Basically the Lyapunov stability analysis
was substituted by the Ricci sectional curvature analysis. In this
report we apply Pettini's geometrical framework to determine the
potential energy of a twisted magnetic flux tube, from its curved
Riemannian geometry. Actually the Lyapunov exponents, are connected
to a Riemann metric tensor, of the twisted magnetic flux tubes
(MFTs). The Hamiltonian flow inside the tube is actually given by
Perelman Ricci flows constraints in twisted magnetic flux tubes,
where the Lyapunov eigenvalue spectra for the Ricci tensor
associated with the Ricci flow equation in MFTs leads to a
finite-time Lyapunov exponential stretching along the toroidal
direction of the tube and a contraction along the radial direction
of the tube. The Jacobi equation for the MFTs is shown to have a
constant sectional Ricci curvature which allows us to compute the
Jacobi-Levi-Civita (JLC) geodesic deviation for the spread of lines
on the tube manifold and chaotic action through the greatest of its
Lyapunov exponents. By analyzing the spectra of the twisted MFT, it
is shown that the greater exponent is positive and proportional to
the random radial flow of the tube, which allows the onset of chaos
is guaranted. The randomness in the twisted flow reminds a discussed
here is similar of a recent work by Shukurov, Stepanov, and Sokoloff
on dynamo action on Moebius flow [\textbf{Phys Rev E 78 (2008)}].
The dynamo action in twisted flux tubes discussed here may also
serve as model for dynamo experiments in laboratory.{\bf PACS
numbers:\hfill\parbox[t]{13.5cm}{02.40.Hw:differential geometries.
91.25.Cw-dynamo theories.}}} \newpage
 \section{Introduction}
 According to Pettini \cite{1}, the geometrization of Hamiltonian chaos can be used as an alternative to Lyapunov exponents \cite{2}
 in the investigation of stability of Hamiltonian flows. In this paper it is shown that not only Hamiltonian enlargement of manifolds can be used in
 this way \cite{1} but also that as shown by Tang and Boozer \cite{3} there is a simple relationship between the metric and the Lyapunov exponents,
 this last approach can be used to investigate the instability of mechanical manifolds \cite{1} through the
 Jacobi-Levi-Civita (JLC) equation case, of simple topologies such as the one of twisted MFTs and even in the analysis of dynamo action \cite{4}.
 Lyapunov spectra of twisted MFTs, are shown to be determined in a simple way, from JLC
 equations, in the case the tube \cite{5} is distorted yielding a negative sectional curvature of $2-D$ manifold $\cal{T}$. The Hamiltonian flow inside the tube is built, and the potential energy V \cite{1},
 is determined by the Ricci scalar. It is shown that, from the computation of Ricci's sectional curvature \cite{1} and the incompressibility condition,
 the sectional curvature is constant in the weak Frenet curvature limit, which allows us to derive a simple form of the JLC geodesic deviation equation, and to solve it.
 The solution appears in terms of the sectional curvature scalar K(s), which in turn in the tube case is proportional to the Frenet curvature ${\kappa}(s)$ of the MFT axis.
 The Lyapunov exponents in this case, are proportional to the Frenet curvature scalar. The JLC vector field $\textbf{J}(s)$, which spread the geodesics along the negative
 sectional curvature manifold. This vector indicate the separation of the geodesics and is built orthogonal velocity $\textbf{v}_{s}(r)$, along the tube axis. Thus
 the constraint $<\textbf{J},\textbf{v}_{s}>=0$, helps one to solve the JLC equation. When the vector $\textbf{J}$ coincides with the radial direction vector
 and is given by the only non-vanishing component $J_{r}:=r(s)$ the JLC equation can be expressed as
 \begin{equation}
 \frac{{d}^{2}J(s)}{ds^{2}}+\frac{1}{2}{R}J(s)=0
 \label{1}
 \end{equation}
where $R=2K_{Ricci}(s)$ is the Ricci curvature scalar, and K
represents the sectional curvature. Here one considers the flux tube
Riemann metric
\begin{equation}
 {d{s_{0}}^{2}}=dr^{2}+r^{2}(s){d{\theta}_{R}}^{2}+K^{2}(s)ds^{2}
 \label{2}
 \end{equation}
 where, $K(s)=(1-r(s){\kappa}(s) \cos{\theta}(s))$ is the stretching
 factor of the metric, and
 ${\theta}(s)={\theta}_{R}-\int{{\tau}(s)ds}$, where ${\tau}(s)$ is
 the Frenet torsion of the flux tube axis, and ${\kappa}(s)$ is the
 Frenet curvature of the same axis. Actually, one considers the thin
 flux tube approximation where $K(s)=1$.
It is well known that when the sectional curvature scalar is
negative the Lyapunov exponents are positive and there is a spread
of the geodesic flow. By analyzing the spectra of the twisted MFT,
it is shown that the greater exponent is proportional to the random
radial flow of the tube, which means that the Riemannian model of
dynamo flow is chaotic. It is also shown that were
 non-decreasing as happens in general with Lyapunov exponents \cite{1}. In this paper one investigates a similar problem in the context of the time evolution
 operator under Ricci flow in the backyard of the twisted MFT. To be able to determine the Lyapunov eigenvalue spectra in twisted MFTs, which
 is a fundamental problem which helps one to determine the stability of the flows inside flux tubes such
 as tokamaks or stellarators in plasma physics \cite{6} or even in
 the context of solar or other stellar plasma loop \cite{6}, one computes the eigenvalue spectra of the
 Ricci tensor under the Ricci flow. This is an important problem for
 plasma theorists and experimentalists, shall be examined here in the
 framework of Riemannian geometry
 \cite{7}. Recently, Thiffeault, Tang and Boozer, investigated Riemannian constraints on Lyapunov exponents \cite{7}, based on the relation between the
 Riemann metric and the finite-time Lyapunov exponential stretching,
 so fundamental for dynamo action. Note that, here, another
 sort of constraint is investigated. Instead of the vanishing of the Riemann curvature tensor, called by mathematicians, Riemann-flat space or condition, one use
 the Lyapunov spectra under Ricci flow. Chaotic flows inside the twisted
 MFTs are investigated. Anosov diffeomorphism \cite{8}, is an important mathematical tool from the theory of
 dynamical systems, that has often been used, in connection with
 the investigation of dynamo flows and maps \cite{9} such as the Arnold's Cat Map \cite{9} on the torus, useful in mixing \cite{10} problems in the
 physics of fluids. One of the main properties of
 the Anosov maps is that they yield Lyapunov exponential of the
 chaotic exponential stretching, which are constant everywhere \cite{7}. The role of the JLC on the Lyapunov exponents on the propagation of light in cosmology
 has been recently addressed by Artyushkova and Sokoloff \cite{11}. After the Perelman's seminal paper \cite{12}, on an explanation of the Poincare conjecture, many applications of the Ricci flow equations to several
 areas of physics have appeared in the literature. This paper is organized as follow: In section 2 the dynamo maps under Ricci
 flows in MFTs are investigated with the aid of Lyapunov spectra. In section 3, the thin tube perturbations are computed in the negative Ricci sectional
 curvature assumption. Section 4 addresses discussions and conclusions.
 \newpage
\section{Ricci fast magnetic dynamo flows in MFTs}
\vspace{1cm} Let us start this section, by defining the Ricci flow
as: \newline \textbf{Definition 2.1}: \newline Let us consider a
smooth manifold which Ricci tensor $\textbf{Ric}$ obeys the
following equation: \begin{equation}
\frac{{\partial}\textbf{g}}{{\partial}t}=2\textbf{Ric} \label{3}
\end{equation}
Here $\textbf{g}$ is the Riemann metric over the manifold $\cal{M}$
where in $\textbf{g}(t)$, $t{\in} [a,b]$. By chosing a local chart
$\cal{U}$ on this manifold, the Ricci flow equation may be written
as
\begin{equation}
\frac{{\partial}{g}_{ij}}{{\partial}t}=-2{R}_{ij} \label{4}
\end{equation}
with this equation in hand let us now compute the eigenvalue spectra
of the Ricci tensor as
\begin{equation}
{R}_{ij}{\chi}^{j}={\lambda}{\chi}_{i} \label{5}
\end{equation}
where here $(i,j=1,2,3)$. Substitution of the Ricci flow equation
(\ref{4}) into the eigenvalue equation (\ref{5}) one obtains an
eigenvalue equation for the metric itself, as
\begin{equation}
\frac{{\partial}{g}_{ij}}{{\partial}t}{\chi}^{j}=-2{\lambda}{g}_{ij}{\chi}^{j}
\label{6}
\end{equation}
and since the eigenvector ${\chi}^{k}$ is in principle arbitrary,
one can reduce this equation to
\begin{equation}
\frac{{\partial}{g}_{ij}}{{\partial}t}=-2{\lambda}{g}_{ij} \label{7}
\end{equation}
which reduces to the solution
\begin{equation}
{g}_{ij}=exp[-2{\lambda}_{i}t]{\delta}_{ij} \label{8}
\end{equation}
one notes that the, ${\delta}_{ij}$ is the Kroenecker delta diagonal
unity matrix. No Einstein sum convention is being used here. By
considering the Tang-Boozer relation between the metric $g_{ij}$
components and the Lyapunov exponents
\begin{equation}
g_{ij}={\Lambda}_{1}\textbf{e}_{1}\textbf{e}_{1}+{\Lambda}_{2}\textbf{e}_{2}\textbf{e}_{2}+{\Lambda}_{3}\textbf{e}_{3}\textbf{e}_{3}
\label{9}
\end{equation}
\newpage
the following lemma, can be proved:
\newline
\textbf{lemma 1}: \newline If ${\lambda}_{i}$ is the eigenvalue
spectra of the Ricci tensor $\textbf{Ric}$ under Ricci flow
\cite{13} equation, the Lyapunov spectra is given by the following
relations:
\begin{equation}
{\lambda}_{i}=-{{\gamma}_{i}}\le{0} \label{10}
\end{equation}
where ${\lambda}_{i}$ are the finite-time Lyapunov numbers. The
infinite or true Lyapunov number is
\begin{equation}
{{\lambda}_{i}}^{\infty}=\lim_{{t\rightarrow\infty}}(\frac{ln{\Lambda}_{i}}{2t})
\label{11}
\end{equation}
Here one has used the finite-time Lyapunov exponent given by
\begin{equation}
{{\lambda}}_{i}=(\frac{ln{\Lambda}_{i}}{2t}) \label{12}
\end{equation}
Note that one of the interesting features of the Ricci flow method
is that one may find the eigenvalue Lyapunov spectra without
computing the Ricci tensor, of the flux tube, for example. Actually
the twisted flux tube Riemannian line element \cite{14}
\begin{equation}
dl^{2}= dr^{2}+r^{2}d{{\theta}_{R}}^{2}+K^{2}(r,s)ds^{2} \label{13}
\end{equation}
can now be used, to compute the Lyapunov exponential stretching of
the flow as in Friedlander and Vishik \cite{14} "dynamo flow" with
the Ricci flow technique above.  Here twist transformation angle is
given by
\begin{equation}
{\theta}(s):={\theta}_{R}-\int{{\tau}(s)ds} \label{14}
\end{equation}
One another advantage of the method used here is that this allows us
to compute the finite-time Lyapunov exponential, without the need of
recurring to the non-Anosov maps \cite{15}. The Lyapunov exponents
here are naturally non-Anosov since the exponents are
non-homogeneous. Here $K(r,s):=(1-r{\kappa}(s,t)\cos{\theta})$, is
the stretching in the metric. Let us now compute the eigenvalue
spectra for the MFTs. Note that the eigenvalue problem, can be
solved by the $3D$ matrix
\begin{equation}
\textbf{M}_{\textbf{3D}}=\pmatrix{2{\lambda}g_{11}&0&0\cr0&{\partial}_{t}g_{22}+{\lambda}g_{22}&0\cr0&0&{\partial}_{t}g_{33}+{\lambda}g_{33}\cr}\qquad
\label{15}
\end{equation}
The eigenvalue equation
\begin{equation}
Det[\textbf{M}_{\textbf{3D}}]=0 \label{16}
\end{equation}
the following eigenvalue Lyapunov spectra for thick tubes, where
$K\approx{-{\kappa}_{0}rcos{\theta}(s)}$ are
\begin{equation}
{\lambda}_{1}=0, {\lambda}_{2}=
2\frac{v_{r}(r)}{r},{\lambda}_{3}=\frac{1}{2}{\lambda}_{2}+{\omega}_{1}\tan{\theta}(s)
\label{17}
\end{equation}
Therefore, since for the existence of dynamo action, at least two of
the Lyapunov exponents have to have opposite signs \cite{1} in order
to obey the stretching $({\lambda}_{3}>0)$ and contracting
$({\lambda}_{2}<0)$, in order that the radial flow ${v}_{r}$ be
negative, ${\lambda}_{3}>0$. Therefore from the above expressions
the following constraint is obtained:
\begin{equation}
|{\omega}_{1}tg{\theta}(s)|\ge{|{\lambda}_{2}|}=|\frac{v_{r}}{r}|
\label{18}
\end{equation}
Thus the $\frac{{v}_{r}}{r}<0$ yields a compression on the flux tube
which induces the in the tube a stretch along the toroidal
direction-s, by the stretch-twist and fold dynamo generation method
of Vainshtein and Zeldovich \cite{17}. In the next section one shall
compute the relation between twist or vorticity, and the when the
sectional curvature of the thin tube is negative. Actually this
leads us to the following time dependence of the magnetic field
components, as
\begin{equation}
{B}_{\theta}\approx{e^{2{\lambda}_{\theta}t}}=e^{(\frac{v_{r}}{r})t}
\label{19}
\end{equation}
\begin{equation}
B_{s}\approx{e^{{\lambda}_{s}t}}=e^{(\frac{v_{r}}{r}+{\omega}_{1}v_{r}tan{\theta})t}
\label{20}
\end{equation}
where ${\omega}_{1}$ is a constant vorticity inside the dynamo flux
tube.

\section{Jacobi-Levi-Civita equation and Ricci curvature}
In this section the Ricci sectional curvature \cite{17} in the case
of a radial perturbation of a thin twisted MFT. This is justified
since, as one has seen in the last section, the radial flow is
fundamental for the existence of non-vanishing Lyapunov exponential
stretching, which in turn are fundamental for the existence of
dynamo action. On the other hand, following work by D. Anosov
\cite{9}, Chicone and Latushkin \cite{18} have previously shown that
geodesic flows, which possesses negative Riemannian curvature has a
fast dynamo action. Let X and Y be vectors laying in tangent
manifolds $\cal{TM}$ to a Riemannian manifold
$\cal{M}\subset{\cal{N}}$ where $\cal{N}$ is an Euclidean three
dimensional space. The Ricci sectional curvature is given by
\begin{equation}
K(X,Y):=\frac{<R(X,Y)Y,X>}{S(X,Y)} \label{21}
\end{equation}
where $R(X,Y)Z$ is the Riemann curvature given by
\begin{equation}
R(X,Y)Z={\nabla}_{X}{\nabla}_{Y}Z-
{\nabla}_{Y}{\nabla}_{X}Z-{\nabla}_{[X,Y]}Z \label{22}
\end{equation}
where
\begin{equation}
S(X,Y):=||{X}||^{2}||Y||^{2}-<X,Y>^{2} \label{23}
\end{equation}
As usual ${\nabla}_{X}Y$ is the Riemannian covariant derivative
given by
\begin{equation}
{\nabla}_{X}Y=({X}.{\nabla})Y \label{24}
\end{equation}
Expression $[X,Y]$ is the commutator, where which on the vector
frame $\textbf{e}_{l}$, $(l=1,2,3)$ in $\textbf{R}^{3}$, as
\begin{equation}
{X}={X}_{k}\textbf{e}_{k}\label{25}
\end{equation}
or its dual basis
\begin{equation}
{X}={X}^{k}{\partial}_{k}\label{26}
\end{equation}
where Einstein summation convention is used. Thus the commutator is
written as
\begin{equation}
[X,Y]={[X,Y]}^{k}{\partial}_{k} \label{27}
\end{equation}
Thus the Riemann curvature tensor becomes
\begin{equation}
R(X,Y)Z=[{R^{l}}_{jkp}Z^{j}X^{k}Y^{p}]{\partial}_{l} \label{28}
\end{equation}
By considering the thin tube approximation $K\approx{1}$ where the
gradient is given by
\begin{equation}
{\nabla}=[{\partial}_{r},r^{-1}{\partial}_{{\theta}_{R}},{\partial}_{s}]
\label{29}
\end{equation}
Note that choosing a chart $U$ $\subset$ $\cal{M}$, one may compute
the Riemann curvature of metric (\ref{2}) as
\begin{equation}
R_{2323}=-\frac{1}{2}r(s)\frac{d^{2}r}{ds^{2}} \label{30}
\end{equation}
which from the JLC equation yields
\begin{equation}
R_{2323}=\frac{1}{2}r^{2}R \label{31}
\end{equation}
It is easy to show from the expression of a isotropic Riemannian
curvature tensor
\begin{equation}
R_{ijkl}=K(g_{ij}g_{kl}-g_{il}g_{jk}) \label{32}
\end{equation}
where $K_{Ricci}$ is the constant sectional curvature, which is
equal to half the Ricci scalar curvature R. Thus from the expression
(\ref{32}) one obtains
\begin{equation}
R_{2323}=K(g_{23}g_{23}-g_{22}g_{33}) \label{33}
\end{equation}
which since the metric is diagonal $g_{23}$ vanishes and this
component of the Riemann tensor coincides with the result
(\ref{32}). The Ricci curvature components are
\begin{equation}
R_{22}=r\frac{d^{2}r}{ds^{2}}=-\frac{1}{2}Rr^{2}\label{34}
\end{equation}
\begin{equation}
R_{33}=\frac{1}{2}R\label{35}
\end{equation}
Now let solve the geodesic deviation JLC equation
\begin{equation}
\frac{{d}^{2}r(s)}{ds^{2}}+K_{Ricci}r(s)=0 \label{36}
\end{equation}
which yields
\begin{equation}
J(s)=\frac{w(s)}{\sqrt{K_{Ricci}}}\sin{\sqrt{(K_{Ricci})s}}
\label{37}
\end{equation}
for $K_{Ricci}>{0}$, and
\begin{equation}
J(s)=s{w(s)} \label{38}
\end{equation}
for $K=0$ and
\begin{equation}
J(s)=\frac{w(s)}{\sqrt{-K_{Ricci}}}\sinh{\sqrt{(-K_{Ricci})s}}
\label{39}
\end{equation}
for $K<0$. Thus if $r=J(s)$ one may say that the tube behaves as a
periodic function in space coordinate-s, in the stable case ($K>0$).
In the stable chaotic case the tube opens and the geodesic particles
of curves diverges. In the case of a MHD flux tube the geodesic
deviation is justified since the charged particles do not follow
geodesics at all, and non-geodetic equations would require a
magnetic Lorentz force on the RHS of the equation such as
\begin{equation}
\frac{d^{2}X^{i}}{ds^{2}}+{{\Gamma}^{i}}_{jk}\frac{dX^{j}}{ds}\frac{dX^{k}}{ds}=0
\label{40}
\end{equation}
where ${{\Gamma}^{i}}_{jk}$ are the Christoffel symbols.
\section{Riemann curvature from Lyapunov exponents
and dynamo action}\vspace{1cm} In this section it is shown that in
regions of weak torsion the Riemann and Ricci curvatures acquire
simple forms, in terms of of exponential stretching and torsion.
Negative torsion in helical chaotic flows, leads to singularities in
curvatures in the $t\rightarrow{\infty}$ time limit. A simple
examination of the Jacobians considered in the last section allows
us to see that the Riemann metric of the flux tubes vanish for a
constant stretch-twist and contraction. However since curvature or
folding is a fundamental process in the STF dynamo mechanism, in
this section it is shown that the consideration of a more general
Jacobian makes the manifold acquire a curved Riemannian metric
distinct from the Euclidean norm metrics considered so far. So, for
a non-uniform stretching \cite{15} usually found in general chaotic
flows the Riemann and Ricci curvatures do not vanish identically.
Let us start computing the curvatures corresponding to the general
Riemann metric of the twisted MFT above. With the aid of an
adaptation of the tensor package the computation of the curvatures
is neither tedious nor long and results in
\begin{equation}
R_{1313}=\frac{1}{4}\frac{{{\tau}_{0}}^{2}cos^{2}({\theta})}{(1-{\tau}_{0}rcos{\theta})}
\label{19}
\end{equation}
\begin{equation}
R_{1323}=-\frac{1}{8}\frac{{{\tau}_{0}}^{2}rsin{2{\theta}}}{(1-{\tau}_{0}rcos{\theta})}
\label{20}
\end{equation}
\begin{equation}
R_{2323}=\frac{1}{4}\frac{{{\tau}_{0}}^{2}r^{2}sin{2{\theta}}}{(1-{\tau}_{0}rcos{\theta})}
\label{21}
\end{equation}
for the Riemann curvature tensor, while the Ricci tensor components
are
\begin{equation}
R_{11}=-\frac{1}{4}\frac{{{\tau}_{0}}^{2}cos^{2}{\theta}}{(1-2{\tau}_{0}rcos{\theta}+{{\tau}_{0}}^{2}rcos{\theta})}
\label{22}
\end{equation}
\begin{equation}
R_{12}=-\frac{1}{8}\frac{{{\tau}_{0}}^{2}rsin{2{\theta}}}{(1-2{\tau}_{0}rcos{\theta}+{{\tau}_{0}}^{2}rcos{\theta})}
\label{23}
\end{equation}
\begin{equation}
R_{33}=-\frac{1}{4}\frac{{{\tau}_{0}}^{2}}{(1-{\tau}_{0}rcos{\theta})}
\label{24}
\end{equation}
\begin{equation}
R_{22}=-\frac{1}{8}\frac{{{\tau}_{0}}^{2}rsin{2{\theta}}}{(1-2{\tau}_{0}rcos{\theta}+{{\tau}_{0}}^{2}rcos{\theta})}
\label{25}
\end{equation}
Most of these curvatures can be easily simplified in the weak
torsion case. Let us now give a more dynamical character to these
expressions by considering the Thiffeault-Boozer Riemann metric
relation with the Lyapunov exponents of exponential stretching by
expressing this metric in terms of the directions $\textbf{t}$,
$\textbf{e}_{r}$ and $\textbf{e}_{\theta}$ along the curved tube as
\begin{equation}
g_{ij}={\Lambda}_{r}\textbf{e}_{r}\textbf{e}_{r}+{\Lambda}_{\theta}\textbf{e}_{\theta}\textbf{e}_{\theta}+{\Lambda}_{s}\textbf{t}\textbf{t}
\label{26}
\end{equation}
where $({i,j=r,{\theta},s})$ and ${\Lambda}_{i}$ are the Lyapunov
numbers which are all positive or null. The Lyapunov exponents are
given by
\begin{equation}
{{\lambda}^{\infty}}_{i}=\lim_{t\rightarrow\infty}(\frac{ln{\Lambda}_{i}}{2t})
\label{27}
\end{equation}
The infinite symbol over the Lyapunov exponent indicates that this
is a true Lyapunov exponent obtained as the limit of the finite-time
Lyapunov exponent ${\lambda}_{i}$. Let us now compute the values of
the Lyapunov exponents which are fundamental for stretching and
dynamos, in terms of a random radial flow
\begin{equation}
{\lambda}_{i}=\lim_{t\rightarrow\infty}(\frac{ln{\Lambda}_{i}}{2t})
\label{28}
\end{equation}
\begin{equation}
<r>=\int{<v_{r}>dt} \label{29}
\end{equation}
where here one shall consider that on a finite-time period, the
random flow can be considered as approximately constant which
reduces the last expression to
\begin{equation}
<r>=<v_{r}>t \label{30}
\end{equation}
Let us now compute the Lyapunov exponents of the curved Riemannian
flux tubes in terms of the random flows as
\begin{equation}
{\lambda}_{r}=\lim_{t\rightarrow\infty}(\frac{ln{1}}{2t})=0
\label{31}
\end{equation}
\begin{equation}
{\lambda}_{\theta}=\lim_{t\rightarrow\infty}(\frac{r}{t})=<v_{r}>
\label{32}
\end{equation}
and finally
\begin{equation}
{\lambda}_{s}=\lim_{t\rightarrow\infty}(\frac{lnK(r,s)}{2t})\approx{\lim_{t\rightarrow\infty}(\frac{-{\tau}_{0}rsin{\theta}}{2t})}\approx{-{\tau}_{0}
<v_{r}>sin{\theta}} \label{33}
\end{equation}
Note that from the above exponents is easy to show that
\begin{equation}
lim(max)[-{\tau}_{0} <v_{r}>sin{\theta}]=[-{\tau}_{0}
<v_{r}>sin{\theta}]> |<v_{r}>|\label{34}
\end{equation}
if one considers the limit of strong torsion. This limit in solar
physics for example, includes the case of kink modes in unstable
solar loops. In this case what happens is that when the tube is
compressed, the magnetic flux tubes possesses a negative radial
random flow $<v_{r}><0$, and the exponent which is negative and
higher exponent becomes positive, which is enough for the onset of
chaos to take place. Following computations by Tang and Boozer
\cite{3}, on the magnetic fields by the self-induction equation one
can say that the corresponding magnetic field components maybe
written in terms of the Lyapunov exponents as
\begin{equation}
{B}_{\theta}\approx{e^{2{\lambda}_{\theta}t}}=e^{<v_{r}>t}
\label{34}
\end{equation}
\begin{equation}
B_{s}\approx{e^{{\lambda}_{s}t}}=e^{-{\tau}_{0}<v_{r}>sin{\theta}t}
\label{35}
\end{equation}
while the radial magnetic field does not depend on time , which
allows us to say that this is due to the confinement of the radial
flow on the tube, actually in the force-free case considered by
Ricca \cite{5} the flux tube radial magnetic component $B_{r}$ is
assumed to vanish. From the last two expressions it is easy to
observe that the magnetic field is stationary in the absence of
random radial flows, and there is no fast dynamo action as well as
no stretching due to the vanishing of the Lyapunov exponents. This
phenomenon is actually the Vishik's anti-fast dynamo theorem
\cite{17} where no fast-dynamo action can be obtained, in
non-stretching flows. Fast dynamo is obtained when the random flow
has a positive average velocity. Note that in this case at least the
$B_{s}$ toroidal field grows in time, while the poloidal field is
spatially periodic and may even grow in time in the case torsion is
negative. To simplify matters let us now compute the Riemann and
Ricci curvatures for the chaotic flow tube metric
\begin{equation}
d{s_{0}}^{2}=dr^{2}+e^{<v_{r}>t}d{\theta}^{2}+e^{-{\tau}_{0}<v_{r}>cos{\theta}t}ds^{2}
\label{36}
\end{equation}
Note that for this expression a positive torsion shows that one of
the Ljapunov exponents in this Riemann metric is positive
(stretching) while the other is negative, representing contraction.
The only non-vanishing component for the Riemann tensor is
\begin{equation}
R_{2323}=-\frac{1}{4}e^{2<v_{r}>{\tau}_{0}(1-cos{\theta})t}sin^{2}{\theta}\label{37}
\end{equation}
which shows that the Riemann curvature is unstable in the infinite
time limit, if the torsion is positive, while it is stable if the
torsion is negative. The Ricci tensor components are
\begin{equation}
R_{33}=e^{(1-{\tau}_{0}cos{\theta})<v_{r}>t}R_{22}\label{38}
\end{equation}
\begin{equation}
R_{22}=\frac{1}{2}cos{\theta}e^{-(1-{\tau}_{0}<v_{r}>t)}\label{39}
\end{equation}
The Ricci tensor is
\begin{equation}
R=\frac{1}{2}e^{-{\lambda}_{\theta}(1-{\tau}_{0})t}sin^{2}{\theta}
\label{40}
\end{equation}
In all these computations one has assumed the weak torsion
approximation, which is very reasoble for example, in astrophysical
plasmas. Note from the Riemann curvature expression that the Riemann
space is flat when the random radial flow vanishes, and when the
torsion ${\tau}_{0}>1$ the scalar curvature R is singular, or
unstable, when $t\rightarrow{\infty}$. However, since one assumes
here that the weak torsion approximation, the curvature computations
are in general unstable. Since the twist of plasma MFTs is
proportional to torsion of the MFT axis, this is a very reasonable
approximation since the twist in kink solar loops for examples is
very weak of the order of $Tw\approx{10^{-10}} cm^{-1}$. When
torsion vanishes the stationary Riemann curvature is periodic and
reduces to
\begin{equation}
R_{2323}=-\frac{1}{4}sin^{2}{\theta}\label{41}
\end{equation}
Thus when the radial random flow vanishes the Riemann curvature can
be spatially periodically. Let us now compute the determinant of the
Riemann metric
\begin{equation}
g=Det{(g_{ij})}=r^{2}(1-{\tau}_{0}rcos{\theta})^{2}\label{42}
\end{equation}
This might be $g=1$ in the case of incompressible chaotic flow as
${\nabla}.\textbf{v}=0$, which allows us to say that the chaotic
helical flow inside the tube has to be compressible. Tang and Boozer
\cite{5}, have also examine the problem of compressible chaotic
flows. As a final observation one notes that the Frenet curvature of
material lines inside the tube, increases as the Ljapunov exponents
or metric decreases. This can be seen by a simple examination of the
metric factor $g_{ss}=(1-{\kappa}(s)rcos{\theta})^{2}$ since the
Frenet curvature ${\kappa}(s)$ growth implies that this stretching
factor decreases, of course along the Ljapunov exponents since they
depend upon the Riemann metric. \newpage
\section{Geodesic flows in
chaotic dynamos} Anosov \cite{8} demonstrated that hyperbolic
systems which are geodesic flows \cite{14}, are necessarilly Anosov.
In this section one shows that as a consequence of the geodesic flow
condition imposed the MFT twisted flows are non-Anosov since their
Lyapunov exponents are not all constants. In this section two given,
the first is the Anosov geodesic flow of Arnold dynamo, while the
second addresses the example of geodesic flows on the twisted MFTs
above. In the first case, the Arnold metric of previous section,
yields the Riemann-Christoffel symbols as
\begin{equation}
{{\Gamma}^{i}}_{jk}=\frac{1}{2}g^{il}(g_{lj,k}+g_{lk,j}-g_{jk,l})\label{43}
\end{equation}
whose components are
\begin{equation}
{{\Gamma}^{1}}_{13}=-\lambda=-{{\Gamma}^{2}}_{23}\label{44}
\end{equation}
\begin{equation}
{{\Gamma}^{3}}_{11}={\lambda}e^{-2{\lambda}z}\label{45}
\end{equation}
\begin{equation}
{{\Gamma}^{3}}_{22}=-{\lambda}e^{2{\lambda}z}\label{46}
\end{equation}
Substitution of these symbols into the geodesic equation
\begin{equation}
\frac{dv^{i}}{dt}+{{\Gamma}^{i}}_{jk}v^{j}v^{k}=0\label{47}
\end{equation}
yields the following equations for the geodesic flow $\textbf{v}$ as
\begin{equation}
\frac{dv^{1}}{dt}+{{\Gamma}^{1}}_{13}v^{1}v^{3}=0\label{48}
\end{equation}
which yields
\begin{equation}
\frac{dv^{1}}{dt}-{\lambda}v^{1}v^{3}=0\label{49}
\end{equation}
Proceeding the same way with the remaining equations yields
\begin{equation}
\frac{dv^{2}}{dt}+{{\lambda}}v^{2}v^{3}=0\label{50}
\end{equation}
and
\begin{equation}
\frac{dv^{3}}{dt}-{\lambda}[e^{-2{\lambda}z}(v^{1})^{2}-e^{2{\lambda}z}(v^{2})^{2}]=0\label{51}
\end{equation}
By assuming that the $v^{3}=constant={v^{3}}_{0}$ one obtains
\begin{equation}
v^{1}=v^{0}e^{{\lambda}[z+{v^{3}}_{0}t]} \label{52}
\end{equation}
\begin{equation}
v^{2}=v^{0}e^{-{\lambda}[z-{v^{3}}_{0}t]}\label{53}
\end{equation}
Since $v^{1}=v^{p}=\frac{dp}{dt}$ and $v^{2}=v^{q}=\frac{dq}{dt}$,
one is able to integrate the geodesic equations to obtain the flow
topology as
\begin{equation}
p^{2}+q^{2}=R^{2}e^{{\lambda}_{L}t}sinh{{\lambda}z}\label{54}
\end{equation}
where ${\lambda}_{L}:=v_{0}{\lambda}$ is the Lyapunov exponent for
the Arnold metric, and $z:=v_{0}t$. This geometry represents a
circle expanding on time and exponentially stretched by the action
of the Lyapunov exponent. Note that the dynamo flow velocity here,
is distinct from the Arnold's one, since the Arnold's dynamo flow
does not possesses components $v^{p}$ and $v^{q}$ as in the example
considered in this section. Let us now proceed computing the same
geodesic flow in the case of twisted MFT metric. Their Christoffel
symbols are
\begin{equation}
{{\Gamma}^{2}}_{33}=\frac{e^{-<v_{r}>(1-{\tau}_{0})t}}{2}sin{\theta}\label{55}
\end{equation}
\begin{equation}
{{\Gamma}^{3}}_{23}=\frac{e^{-<v_{r}>(1-{\tau}_{0}t}}{2}sin{\theta}\label{56}
\end{equation}
From these expressions one obtains the following geodesic equations
\begin{equation}
\frac{dv^{1}}{dt}=0\label{57}
\end{equation}
which implies $v^{r}=v^{1}=v^{0}=constant$. This result is indeed
important since, it shows that actually our hypothesis that the
random radial flow is constant is feasible, and can be derived from
a geodesic flow. To simplify the remaining equations one chooses a
torsion ${\tau}_{0}=\frac{1}{2}$ , which yields
\begin{equation}
\frac{dv^{2}}{dt}-sin{\theta}(v^{3})^{2}=0\label{58}
\end{equation}
and
\begin{equation}
\frac{dv^{3}}{dt}-e^{[1-\frac{1}{2}<v_{r}>t]}{v^{2}}{v^{3}}=0\label{59}
\end{equation}
These two last equations together yield
\begin{equation}
\frac{dv^{3}}{dv^{2}}=e^{-\frac{1}{2}<v_{r}>t}\frac{v^{2}}{v^{3}}\label{60}
\end{equation}
Some algebra yields the following result
\begin{equation}
\frac{v_{s}}{v^{\theta}}=e^{(1+cos{\theta})<v_{r}>t}\label{61}
\end{equation}
Thus the toroidal flow is proportional to the random flow and to the
Lyapunov exponential stretching. Note that at $t=0$ there is an
equipartition between the poloidal and toroidal flows while as time
evolves the Lyapunov exponential stretching makes the toroidal flow
velocity to increase without bounds with respect to the poloidal
flow. This is actually analogous to the dynamo action. Equation
(\ref{57}) also shows that one the Lyapunov exponents is constant
while the other is not and therefore the flow is a non-Anosov flow.

\section{Conclusions}
Let us state here for discussion the Chicone-Latushkin fast dynamo
in geodesic flow theorem \cite{17} as:
\newline
\textbf{Theorem}: If $\textbf{v}$ is the vector field that generates
the geodesic flow for a closed two dimensional Riemannian manifold
$\cal{M}$ . then $\textbf{v}$ is a steady solution of Euler´s
equation on $\cal{M}$. In addition if $\cal{M}$ has constant
negative curvature $\kappa$, then for each magnetic Reynolds number
$Re_{m}>\sqrt{-{\kappa}}$, the corresponding dynamo operator has a
positive eigenvalue given by ${\lambda}_{\epsilon}$ above.
\newline
Therefore one may say that in this paper we give an example of the
validity of Chicone-Latushkin theorem on geodesic dynamo flows in
the case of flux tubes where a containing Ricci plasma flow. Also in
this paper the importance of investigation the Ricci flows as a
constraint to dynamo flows inside twisted MFTs is stressed, given
examples of the dynamo action existence in negative sectional
curvature. This can be basically done with the help of the Lyapunov
spectra of the Ricci fast dynamo flows, which are fundamental for
the exponenial stretching which are in turn so important for dynamo
action. An instability of the Lyapunov exponential stretching
influence on the instability of the Euler equations has been
discussed by Friedlander and Vishik \cite{17}. In the case discussed
above it is shown that compressed solar flux tubes for example can
give rise to the onset of chaos yielding chaotic dynamos \cite{18}
to take place. Further investigation between the relation between
the random flows in Riemannian space and Lyapunov exponents
\cite{19,20} may appear elsewhere. A very recently work on dynamo
action in Moebius flows \cite{21} and its connection with the also
twisted flow discussed here may appear elsewhere. Experiments for
dynamo action in laboratory may also be suggested from the work
discussed here.
\section{Acknowledgements}:
\newline I also am deeply indebt to G. Paternain , Dmitry Sokoloff and R Ricca, for helpful discussions on the subject of this
paper. Financial supports from Universidade do Estado do Rio de
Janeiro (UERJ) and CNPq (Brazilian Ministry of Science and
Technology) are highly appreciated.


\begin{thebibliography}{20}
\bibitem{12} G. Perelman, The entropy formula for the Ricci flow and its geometric applications, Los Alamos Arxives: math/0211599.
\end{thebibliography}
 \end{document}